\documentclass[acmsmall,nonacm=true,screen=true]{acmart}

\AtBeginDocument{%
  \providecommand\BibTeX{{%
    \normalfont B\kern-0.5em{\scshape i\kern-0.25em b}\kern-0.8em\TeX}}}

\setcopyright{acmcopyright}
\copyrightyear{2022}
\acmYear{2022}
\acmDOI{10.1145/123123123.121212}

\acmConference[ICMLC '22]{ICMLC '22}{February 18--21, 2022}{Guangzhou, China}
\acmBooktitle{ICMLC '22, February 18--21, 2022, Guangzhou, China}
\acmPrice{15.00}
\acmISBN{978-1-4503-XXXX-X/22/02}

\usepackage{subfig}

\usepackage{tikz}
\usetikzlibrary{positioning, calc, shapes.geometric, shapes.symbols, arrows.meta, fit}

\usepackage{pgfplotstable}
\pgfplotsset{compat=1.16}

\usepackage{lmodern}
\begin{document}

\title{AI for Closed-Loop Control Systems}
\subtitle{New Opportunities for Modeling, Designing, and Tuning Control Systems}

\author{Julius Schöning}
\email{j.schoening@hs-osnabrueck.de}
\affiliation{%
  \institution{Osnabrück University of Applied Sciences}
  \department{Faculty of Engineering and Computer Science}
  \city{Osnabrück}
  \country{Germany}
  \postcode{DE-49076}
}

\author{Adrian Riechmann}
\email{adrian.riechmann@kea-nds.de}

\affiliation{%
  \institution{Science to Business GmbH}
  \department{Competence Centre of Electronics and Electric Drive Systems}
  \city{Osnabrück}
  \country{Germany}
  \postcode{DE-49076}
}

\author{Hans-Jürgen Pfisterer}
\email{j.pfisterer@hs-osnabrueck.de}

\affiliation{%
  \institution{Osnabrück University of Applied Sciences}
  \department{Faculty of Engineering and Computer Science}
  \city{Osnabrück}
  \country{Germany}
  \postcode{DE-49076}
}

\renewcommand{\shortauthors}{}
\renewcommand{\shorttitle}{}

\begin{abstract}
  Control Systems, particularly closed-loop control systems (CLCS), are frequently used in production machines, vehicles, and robots nowadays. CLCS are needed to actively align actual values of a process to a given reference or set values in real-time with a very high precession. Yet, artificial intelligence (AI) is not used to model, design, optimize, and tune CLCS. This paper will highlight potential AI-empowered and -based control system designs and designing procedures, gathering new opportunities and research direction in the field of control system engineering. Therefore, this paper illustrates which building blocks within the standard block diagram of CLCS can be replaced by AI, i.e., artificial neuronal networks (ANN). Having processes with real-time contains and functional safety in mind, it is discussed if AI-based controller blocks can cope with these demands. By concluding the paper, the pros and cons of AI-empowered as well as -based CLCS designs are discussed, and possible research directions for introducing AI in the domain of control system engineering are given.
\end{abstract}




\maketitle

\section{Introduction}
In an ideal world, the controller of CLCS sets the control variables of processes or systems so that the system response perfectly matches the exact reference value. The intended process response is strictly adhered to the reference value---at every time, no matter how high the disturbance on the process is. However, real-world processes and systems such as mechatronical acutuators within vehicles, bio-chemical processes within reactors, and temperature regulation systems within buildings can not be controlled without any over as well as under shootings, settling times, and oscillations. Thus, every technology bringing real-world controllers closer to the theoretical ideal controllers is welcome. Admittedly, even with nowadays computational power, the general procedure of designing control systems \cite{Skogestad2005,Chin2017} has not changed within the last four decades.

Introducing AI in the domain of closed-loop and feedback control systems, hereinafter referred to as CLCS, will improve the entire CLCS and the controller behavior, especially for multivariable and nonlinear processes. Further AI might be capable of predicting possible disturbances even before they affect the process. It is conceivable that AI will replace the conventional controller block entirely and directly interpret the sensor values as process feedback. As a result, the CLCS block diagrams and the controller design procedure will adopt, enabling AI within control system engineering.

In the remainder of this paper, the state of the art concerning CLCS and physics-informed AI are summarized in Section \ref{sec:stoa}. Discussing new AI-empowered and -based block diagrams for CLCS, Section \ref{sec:aiCLCS} will introduce new blocks and new design patterns in detail. Focusing on functional safety, the most critical property of CLCS for specific processes, Section \ref{sec:safeCLCS} will show that AI-based CLCS and safety will go together. By concluding this paper, Section \ref{sec:futherSteps} the pros and cons of AI-empowered as well as -based CLCS designs are discussed and possible research directions for the usage of AI within CLCS are highlighted. In this paper, for simplification, the term process stands for both: the controlled process and the controlled systems. Whereas the term system expresses the entire open- or closed-loop system, hereafter.

\section{State of the Art}\label{sec:stoa}
\begin{figure}[htb!]
	\centering

  \pgfplotstableread[col sep = comma]{images/literatureCount.csv}\datatable

\subfloat[publication count ${\left[ 0,40,000 \right]}$ ] {
  \resizebox{\linewidth}{!}{
  \begin{tikzpicture}
  \begin{axis}[
      width=24cm,
      height=8cm,
      ymin=0,
      ymax=40000,
      xmin=1961,
      xmax=2021,
      xtick={1961,1971,...,2021},
      xmajorgrids,
      ymajorgrids,
      legend pos=north west,
      xticklabel style={/pgf/number format/.cd,
                        1000 sep = {}},
  ]
  \addplot [red, mark = * ] table [x ={YEAR}, y={ANN}]{\datatable};
    \addlegendentry{Artificial Neural Network}
  \addplot [green, mark = *  ] table [x ={YEAR}, y={AI}]{\datatable};
      \addlegendentry{Artificial Intelligence}
  \addplot [blue, mark = * ] table [x ={YEAR}, y={CLCS}]{\datatable};
      \addlegendentry{Closed-Loop Control Systems}
  \addplot [magenta, mark = * ] table [x ={YEAR}, y={CLCSANN}]{\datatable};
      \addlegendentry{Closed Loop Control Systems \& Neural Network}
  \addplot [cyan, mark = * ] table [x ={YEAR}, y={CLCSAI}]{\datatable};
          \addlegendentry{Closed Loop Control Systems \& Artificial Intelligence}
  \end{axis}

\end{tikzpicture}}}

\subfloat[publication count ${\left[ 0,100 \right]}$ ] {
\resizebox{\linewidth}{!}{
\begin{tikzpicture}
\begin{axis}[
    width=24cm,
    height=8cm,
    ymin=0,
    ymax=100,
    xmin=1961,
    xmax=2021,
    xtick={1961,1971,...,2021},
    xmajorgrids,
    ymajorgrids,
    legend pos=north west,
    xticklabel style={/pgf/number format/.cd,
                      1000 sep = {}},
]
\addplot [magenta, mark = * ] table [x ={YEAR}, y={CLCSANN}]{\datatable};
    \addlegendentry{Closed Loop Control Systems \& Neural Network}
\addplot [cyan, mark = * ] table [x ={YEAR}, y={CLCSAI}]{\datatable};
        \addlegendentry{Closed Loop Control Systems \& Artificial Intelligence}
\end{axis}

\end{tikzpicture}}}

\caption{Count of publication with the keywords: \textit{Closed-Loop Control Systems}, \textit{Artificial Neuronal Network}, \textit{Artificial Intelligence} and the keyword combinations: \textit{Closed-Loop Control Systems \& Neuronal Network}, \textit{Closed-Loop Control Systems \& Artificial Intelligent}; calculated from the Elservier Scopus bibliographic database \cite{Elsevier2020}.
}
\label{fig:keywords}
\end{figure}
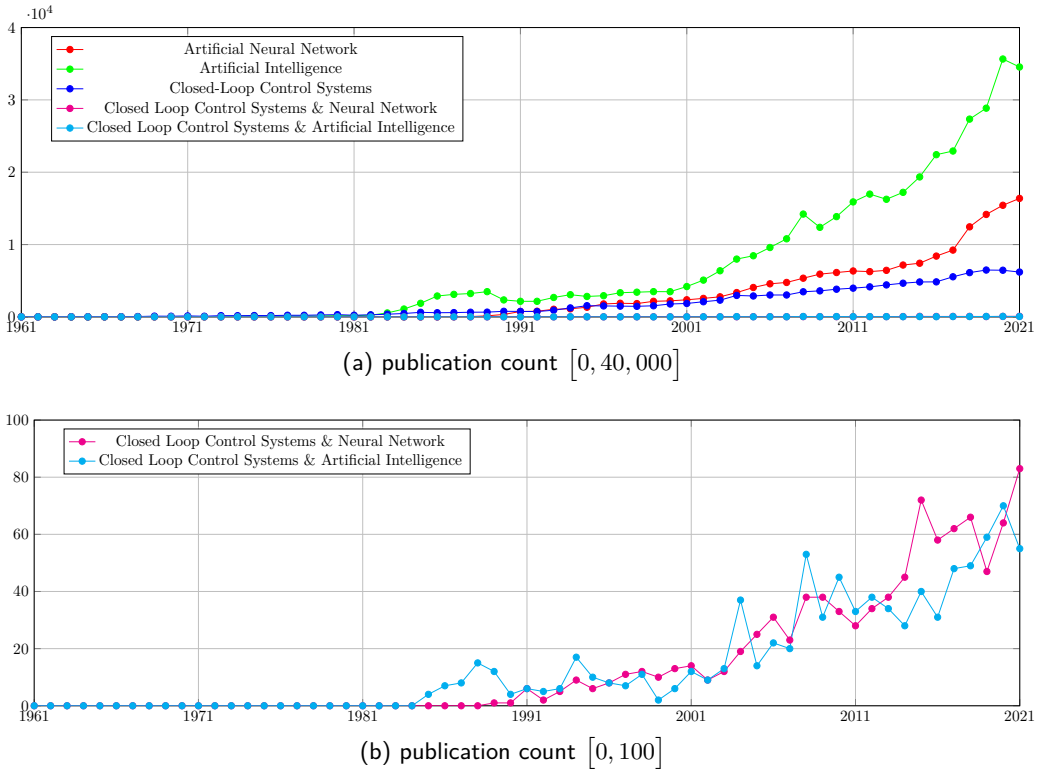

AI and especially ANN have not yet made significant inroads into the field of control systems, including both open- as well as closed-loop systems. Figure \ref{fig:keywords} prove this assessment by comparing the different curve gradients of the publication count with the keyword \textit{Neural Network} in contrast to the publication count with the keyword combination \textit{Closed-Loop Control Systems} and \textit{Neural Network}. Some reasons for the, so far, non-existing combination of control system engineering and AI can be the following:
\begin{itemize}
  \item the complexity of the phase space described by the differential equations,
  \item the unpredictability of the disturbances on the process to be controlled, as well as
  \item the tradition in control system engineering, which is characterized by describing the process in its entirety in physical models.
\end{itemize}

\subsection{Classical Closed Loop Control Systems}
In CLCS, the most common controller used is the proportional– integral–derivative (PID) controller. This high usage is because most engineers understand how to vary the three parameters \cite{Atherton1999}, based on the past (I), present (P), and future (D) control error \cite{Astroem2001} for achieving the intended CLCS behavior. However, PID controllers do not work well for nonlinear processes as well as for processes with small dead-times \cite{Atherton1999}. Further, depending on its parameter, PID controllers are prone to over- and undershot or reach the set-point very slow \cite{Sung1996}. As countermeasures, the parameter turning strategies \cite{Sung1996} have been optimized, and the single PID controller design has been decomposition more complex designs such as  PID-P, PI-PD, and PI-D controller \cite{Atherton1999}.

Nonlinear processes with uncertainties distributed non-periodic values are still hard to control \cite{Astroem2001}. 
Processes with multivariable responses and reference are also unfeasible to control by conventional CLCS design. 

\subsection{AI-Based Closed Loop Control Systems}
The gap between the non-existing AI and control system engineering combination was recently recognized, so the first scientific papers appeared. For example, Rackauckas et al. \cite{Rackauckas2021} were recently able to present a general approach using ANN to approximate differential equations, even finding missing terms. The developed methodology of universal differential equations discovery can approximate the differential equations of a specific process based on its recorded data.

For combining AI and CLCS, differentiable programming languages like Julia \cite{Bezanson2017} are needed to solve the differential equations and train the ANN within the same software code. The benefit of Julia as a programming language is shown inter alia by Rackauckas et al. \cite{Rackauckas2020}. The concept of physics-informed neural networks \cite{Raissi2019} is based on the same conceptual idea but uses different programming languages to bring differential equations and ANN together.
Still to be mentioned are the two more review papers on nonlinear \cite{Raissi2020} and dynamic \cite{Brunton2016} differential equations and ANN, respectively. In engineering applications, the gap between control engineering and AI currently continues to exist.

\subsection{Procedure of the Control System Design}
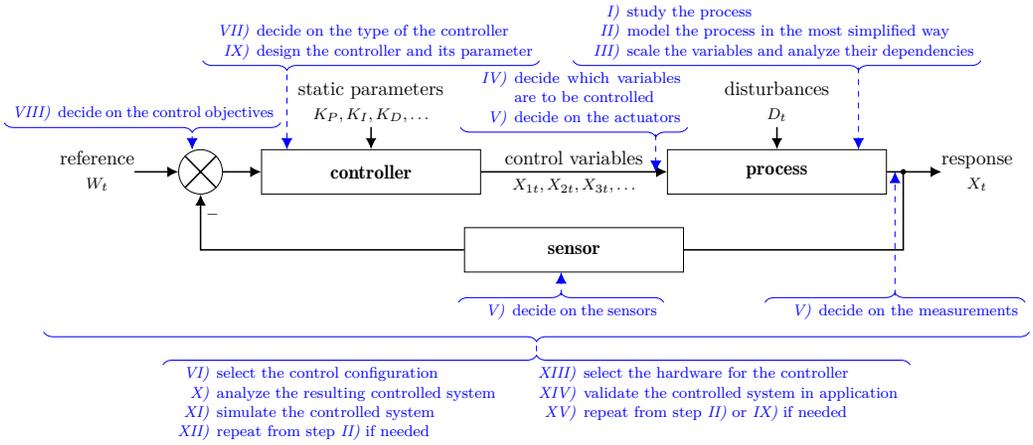
\begin{figure}[htb!]
	\centering
\resizebox{\linewidth}{!}
{
\begin{tikzpicture}
  \node(controller)[draw, rectangle, minimum width = 4.0cm, minimum height = 0.8cm] {\textbf{controller}};

  \node(parameter)[above = 0.3cm of controller, rectangle, minimum height = 0.8cm] {\begin{tabular}{c}
  static parameters \\ \footnotesize $(K_{P},K_{I},K_{D},\ldots)$
  \end{tabular}
  };

  \node(process)[draw, right = 3.4cm of controller, rectangle, minimum width = 4.0cm, minimum height = 0.8cm] {\textbf{process}};

  \node(disturbances)[above = 0.3cm of process, rectangle, minimum height = 0.8cm] {\begin{tabular}{c}
  disturbances \\ \footnotesize $(D_{(t)})$
  \end{tabular}
  };

  \node(sensor)[draw, below right = 1cm and 1.7cm of controller, rectangle, minimum width = 4.0cm, minimum height = 0.8cm, anchor=center] {\textbf{sensor}};

  \node(response)[right = 0.7cm of process, rectangle, minimum height = 0.8cm] {\begin{tabular}{c}
   response \\ \footnotesize $(X_{(t)})$
  \end{tabular}
  };
  \node(sum)[draw, left = 0.7cm of controller, circle, minimum height = 0.8cm] {
  };
  \draw [thick] (sum.north east) -- (sum.south west)
  (sum.north west) -- (sum.south east);

  \node(reference)[left = 0.5cm of sum, rectangle, minimum height = 0.8cm] {\begin{tabular}{c}
   reference \\ \footnotesize $(W_{(t)})$
  \end{tabular}
  };

  \draw [-{Latex[length=2mm, width=2mm]}, thick, draw=black](controller.east)--(process.west) node [above,pos=0.5] (controlVariables) {control variables} node [below,pos=0.5] {\footnotesize $(X_{1(t)},X_{2(t)},X_{3(t)},\ldots)$};
  \draw [-{Latex[length=2mm, width=2mm]}, thick, draw=black](process.east)--($(response.west)+(0.3,0)$) node (rs) [circle, pos=0.3, fill=black, inner sep=0pt,minimum size=1mm] {};
  \draw [-{Latex[length=2mm, width=2mm]}, thick, draw=black](sum.east)--(controller.west);
  \draw [-{Latex[length=2mm, width=2mm]}, thick, draw=black]($(reference.east)+(-0.3,0)$)--(sum.west);
  \draw [-{Latex[length=2mm, width=2mm]}, thick, draw=black]($(disturbances.south)+(0,0.1)$)--(process.north);
  \draw [-{Latex[length=2mm, width=2mm]}, thick, draw=black]($(parameter.south)+(0,0.1)$)--(controller.north);
  \draw [thick, draw=black](rs.center)|-(sensor.east);
  \draw [-{Latex[length=2mm, width=2mm]}, thick, draw=black](sensor.west)-|(sum.south) node [right,pos=0.8] {\textbf{--}};

  \node(processText)[above = 1.5cm of process, rectangle] {\color{blue}\footnotesize{\begin{minipage}{7.2cm}\begin{itemize}
      \item[\textit{I)}] study the process
      \item[\textit{II)}] model the process in the most simplified way
      \item[\textit{III)}] scale the variables and analyze their dependencies
\end{itemize}\end{minipage}}
  };
  \draw[decorate,decoration={brace,amplitude=5pt, aspect=0.3}, draw=blue] ($(processText.south east)+(0,0.1)$) -- node[below=0pt, pos=0.3] (processCBrace) {}  ($(processText.south west)+(0,0.1)$);

  \draw [-{Latex[length=2mm, width=2mm]}, dashed, draw=blue] let
  \p1=($(process.north)$), \p2=($(processCBrace.center)$) in (processCBrace.center)--(\x2,\y1);

  \node(controlText)[above = 0.2cm of controlVariables, rectangle] {\color{blue}\footnotesize{\begin{minipage}{3.9cm}\begin{itemize}
      \item[\textit{IV)}] decide which variables are to be controlled
      \item[\textit{V)}] decide on the actuators
\end{itemize}\end{minipage}}
  };
  \draw[decorate,decoration={brace,amplitude=5pt, aspect=0.14}, draw=blue] ($(controlText.south east)+(0,0.1)$) -- node[below=0pt, pos=0.14] (controlCBrace) {}  ($(controlText.south west)+(0,0.1)$);

  \draw [-{Latex[length=2mm, width=2mm]}, dashed, draw=blue] let
  \p1=($(process.east)$), \p2=($(controlCBrace.center)$) in (controlCBrace.center)--(\x2,\y1);

  \node(measurementsText)[below right = 1.85cm and -2.225cm of process, rectangle] {\color{blue}\footnotesize{\begin{minipage}{4.5cm}\begin{itemize}
      \item[\textit{V)}] decide on the measurements
\end{itemize}\end{minipage}}
  };
  \draw[decorate,decoration={brace,amplitude=5pt, aspect=0.5}, draw=blue] ($(measurementsText.north west)-(0,0.1)$) -- node[above=0pt, pos=0.5] (measurementsCBrace) {}  ($(measurementsText.north east)-(0,0.1)$);

  \draw [-{Latex[length=2mm, width=2mm]}, dashed, draw=blue] let
  \p1=($(process.west)$), \p2=($(measurementsCBrace.center)$) in (measurementsCBrace.center)--(\x2,\y1);

  \node(sensorText)[left = 1.85cm of measurementsText, rectangle] {\color{blue}\footnotesize{\begin{minipage}{3.5cm}\begin{itemize}
      \item[\textit{V)}] decide on the sensors
  \end{itemize}\end{minipage}}
  };
  \draw[decorate,decoration={brace,amplitude=5pt, aspect=0.5}, draw=blue] ($(sensorText.north west)-(0,0.1)$) -- node[above=0pt, pos=0.5] (sensorCBrace) {}  ($(sensorText.north east)-(0,0.1)$);

  \draw [-{Latex[length=2mm, width=2mm]}, dashed, draw=blue] let
  \p1=($(sensor.south)$), \p2=($(sensorCBrace.center)$) in (sensorCBrace.center)--(\x2,\y1);

  \node(controllerText)[above = 1.5cm of controller, rectangle] {\color{blue}\footnotesize{\begin{minipage}{5.9cm}\begin{itemize}
      \item[\textit{VII)}] decide on the type of the controller
      \item[\textit{IX)}] design the controller and its parameter
  \end{itemize}\end{minipage}}
  };
  \draw[decorate,decoration={brace,amplitude=5pt, aspect=0.75}, draw=blue] ($(controllerText.south east)+(0,0.1)$) -- node[below=0pt, pos=0.75] (controlerCBrace) {}  ($(controllerText.south west)+(0,0.1)$);

  \draw [-{Latex[length=2mm, width=2mm]}, dashed, draw=blue] let
  \p1=($(controller.north)$), \p2=($(controlerCBrace)$) in (controlerCBrace)--(\x2,\y1);

  \node(sumText)[above left = 0.5cm and -1.75cm of sum, rectangle] {\color{blue}\footnotesize{\begin{minipage}{4.8cm}\begin{itemize}
      \item[\textit{VIII)}] 	decide on the control objectives
  \end{itemize}\end{minipage}}
  };
  \draw[decorate,decoration={brace,amplitude=5pt, aspect=0.32}, draw=blue] ($(sumText.south east)+(0,0.1)$) -- node[below=0pt, pos=0.32] (sumCBrace) {}  ($(sumText.south west)+(0,0.1)$);

  \draw [-{Latex[length=2mm, width=2mm]}, dashed, draw=blue] let
  \p1=($(sum.north)$), \p2=($(sumCBrace)$) in (sumCBrace)--(\x2,\y1);

  \draw[decorate,decoration={brace,amplitude=5pt, aspect=0.5}, draw=blue] let
  \p1=($(reference.west)$), \p2=($(response.east)$), \p3=($(sensorText.south)$) in ($(\x2,\y3)-(0,0.1)$) -- node[below=0pt, pos=0.5] (systemCBrace) {}  ($(\x1,\y3)-(0,0.1)$);

  \node(systemText)[below = 0.25cm of systemCBrace, rectangle] {\color{blue}\footnotesize{\begin{minipage}[t]{6.6cm}\begin{itemize}
      \item[\textit{VI)}] select the control configuration
      \item[\textit{X)}] analyze the resulting controlled system
      \item[\textit{XI)}] simulate the controlled system
      \item[\textit{XII)}] repeat from step \textit{II)} if needed
  \end{itemize}\end{minipage}
  \begin{minipage}[t]{6.6cm}\begin{itemize}
      \item[\textit{XIII)}] select the hardware for the controller
      \item[\textit{XIV)}] validate the controlled system in application
      \item[\textit{XV)}] repeat from step \textit{II)} or \textit{IX)} if needed
  \end{itemize}\end{minipage} }
  };
  \draw[decorate,decoration={brace,amplitude=5pt, aspect=0.5}, draw=blue] ($(systemText.north west)-(0,0.1)$) -- node[above=0pt, pos=0.5] (systemIICBrace) {}  ($(systemText.north east)-(0,0.1)$);

  \draw [dashed, draw=blue] (systemCBrace)--(systemIICBrace);

  \end{tikzpicture}
}
\caption{The process of control system design, first 14 steps of Skogestad and Postlethwaite \cite{Skogestad2005} in accordance visualized with the corresponding blocks of the CLCS block diagram.}
\label{fig:CLS_process}
\end{figure}

Starting at the turn of the millennium, an increasing amount of computer-aided applied control designing software solutions like Simulink (MATLAB) \cite{MathWorks2021},  Xcos (Scilab) \cite{ESIGroup2021}, and control (Octave) \cite{OFC2021} occurs and manifests the currently used best practices in the procedure of the controller design. As illustrated in Figure \ref{fig:CLS_process}, 14 steps are common to get from the process needed to control to the final control system design \cite{Skogestad2005}. Note, in Figure \ref{fig:CLS_process} an additional 15. step is added due to personal experience.

\section{AI empowerment for Closed-Loop Control Systems}\label{sec:aiCLCS}
Considering the control system design procedure and the CLCS block diagram, AI can be applied in different design steps and blocks. Thus, this Section will first introduce AI for studying, modeling, and scaling the real-world process which needs to be controlled. Thereafter the benefits of AI-based controller tuning and entirely AI-based controller are discussed. Finally, the resulting changes in the control system design procedure, raised by the introduction of AI in control system engineering, are recapitulated.

\begin{figure}[htb!]
	\centering
	\subfloat[hand-crafted physical process model \label{fig:CLS_phy}] {
  \resizebox{\linewidth}{!}
  {
  \begin{tikzpicture}
  \node(controller)[draw, rectangle, minimum width = 4.0cm, minimum height = 0.8cm] {\textbf{controller}};

  \node(parameter)[above = 0.3cm of controller, rectangle, minimum height = 0.8cm] {\begin{tabular}{c}
  static parameters \\ \footnotesize $(K_{P},K_{I},K_{D},\ldots)$
  \end{tabular}
  };

  \node(process)[draw, right = 3.4cm of controller, rectangle, minimum width = 4.0cm, minimum height = 0.8cm] {\textbf{process}};

  \node(disturbances)[above = 0.3cm of process, rectangle, minimum height = 0.8cm] {\begin{tabular}{c}
  disturbances \\ \footnotesize $(D_{(t)})$
  \end{tabular}
  };

  \node(sensor)[draw, below right = 1cm and 1.7cm of controller, rectangle, minimum width = 4.0cm, minimum height = 0.8cm, anchor=center] {\textbf{sensor}};

  \node(response)[right = 0.7cm of process, rectangle, minimum height = 0.8cm] {\begin{tabular}{c}
   response \\ \footnotesize $(X_{(t)})$
  \end{tabular}
  };
  \node(sum)[draw, left = 0.7cm of controller, circle, minimum height = 0.8cm] {
  };
  \draw [thick] (sum.north east) -- (sum.south west)
  (sum.north west) -- (sum.south east);

  \node(reference)[left = 0.5cm of sum, rectangle, minimum height = 0.8cm] {\begin{tabular}{c}
   reference \\ \footnotesize $(W_{(t)})$
  \end{tabular}
  };

  \draw [-{Latex[length=2mm, width=2mm]}, thick, draw=black](controller.east)--(process.west) node [above,pos=0.5] (controlVariables) {control variables} node [below,pos=0.5] {\footnotesize $(X_{1(t)},X_{2(t)},X_{3(t)},\ldots)$};
  \draw [-{Latex[length=2mm, width=2mm]}, thick, draw=black](process.east)--($(response.west)+(0.3,0)$) node (rs) [circle, pos=0.3, fill=black, inner sep=0pt,minimum size=1mm] {};
  \draw [-{Latex[length=2mm, width=2mm]}, thick, draw=black](sum.east)--(controller.west);
  \draw [-{Latex[length=2mm, width=2mm]}, thick, draw=black]($(reference.east)+(-0.3,0)$)--(sum.west);
  \draw [-{Latex[length=2mm, width=2mm]}, thick, draw=black]($(disturbances.south)+(0,0.1)$)--(process.north);
  \draw [-{Latex[length=2mm, width=2mm]}, thick, draw=black]($(parameter.south)+(0,0.1)$)--(controller.north);
  \draw [thick, draw=black](rs.center)|-(sensor.east);
  \draw [-{Latex[length=2mm, width=2mm]}, thick, draw=black](sensor.west)-|(sum.south) node [right,pos=0.8] {\textbf{--}};

  \end{tikzpicture}
  }
  }

  \subfloat[AI generated empirical process model \label{fig:CLS_emp}] {
  \resizebox{\linewidth}{!}
  {
  \begin{tikzpicture}
  \node(controller)[draw, rectangle, minimum width = 4.0cm, minimum height = 0.8cm] {\textbf{controller}};

  \node(parameter)[above = 0.3cm of controller, rectangle, minimum height = 0.8cm] {\begin{tabular}{c}
  static parameters \\ \footnotesize $(K_{P},K_{I},K_{D},\ldots)$
  \end{tabular}
  };

  \node(process)[draw=red, ultra thick, right = 3.4cm of controller, rectangle, minimum width = 4.0cm, minimum height = 0.8cm] {\color{red}\textbf{process}};

  \node(disturbances)[above = 0.3cm of process, rectangle, minimum height = 0.8cm] {\begin{tabular}{c}
  disturbances \\ \footnotesize $(D_{(t)})$
  \end{tabular}
  };

  \node(sensor)[draw, below right = 1cm and 1.7cm of controller, rectangle, minimum width = 4.0cm, minimum height = 0.8cm, anchor=center] {\textbf{sensor}};

  \node(response)[right = 0.7cm of process, rectangle, minimum height = 0.8cm] {\begin{tabular}{c}
   response \\ \footnotesize $(X_{(t)})$
  \end{tabular}
  };
  \node(sum)[draw, left = 0.7cm of controller, circle, minimum height = 0.8cm] {
  };
  \draw [thick] (sum.north east) -- (sum.south west)
  (sum.north west) -- (sum.south east);

  \node(reference)[left = 0.5cm of sum, rectangle, minimum height = 0.8cm] {\begin{tabular}{c}
   reference \\ \footnotesize $(W_{(t)})$
  \end{tabular}
  };

  \draw [-{Latex[length=2mm, width=2mm]}, thick, draw=black](controller.east)--(process.west) node [above,pos=0.5] (controlVariables) {control variables} node [below,pos=0.5] {\footnotesize $(X_{1(t)},X_{2(t)},X_{3(t)},\ldots)$};
  \draw [-{Latex[length=2mm, width=2mm]}, thick, draw=black](process.east)--($(response.west)+(0.3,0)$) node (rs) [circle, pos=0.3, fill=black, inner sep=0pt,minimum size=1mm] {};
  \draw [-{Latex[length=2mm, width=2mm]}, thick, draw=black](sum.east)--(controller.west);
  \draw [-{Latex[length=2mm, width=2mm]}, thick, draw=black]($(reference.east)+(-0.3,0)$)--(sum.west);
  \draw [-{Latex[length=2mm, width=2mm]}, thick, draw=black]($(disturbances.south)+(0,0.1)$)--(process.north);
  \draw [-{Latex[length=2mm, width=2mm]}, thick, draw=black]($(parameter.south)+(0,0.1)$)--(controller.north);
  \draw [thick, draw=black](rs.center)|-(sensor.east);
  \draw [-{Latex[length=2mm, width=2mm]}, thick, draw=black](sensor.west)-|(sum.south) node [right,pos=0.8] {\textbf{--}};

  \node(recordedData)[draw=red, ultra thick, above left = 2cm and 1.0cm of process, rectangle, minimum width = 3.0cm, minimum height = 0.8cm] {\color{red}{\begin{tabular}{c}\textbf{recorded process}\\ \textbf{data}
  \end{tabular}
  }};

  \draw [-{Latex[length=2mm, width=2mm]}, draw=red, ultra thick] (recordedData.east)-|($(process.north)-(1.25,0)$);

  \end{tikzpicture}
  }
  }

\caption{CLCS block diagrams, \protect\subref{fig:CLS_phy} with a hand-crafted physical process model and \protect\subref{fig:CLS_emp} with a empirical process model, e.g. trained by an ANN on recorded data; both process models meet the needs during controller design procedure.}
\label{fig:CLS_pyh_emp}
\end{figure}
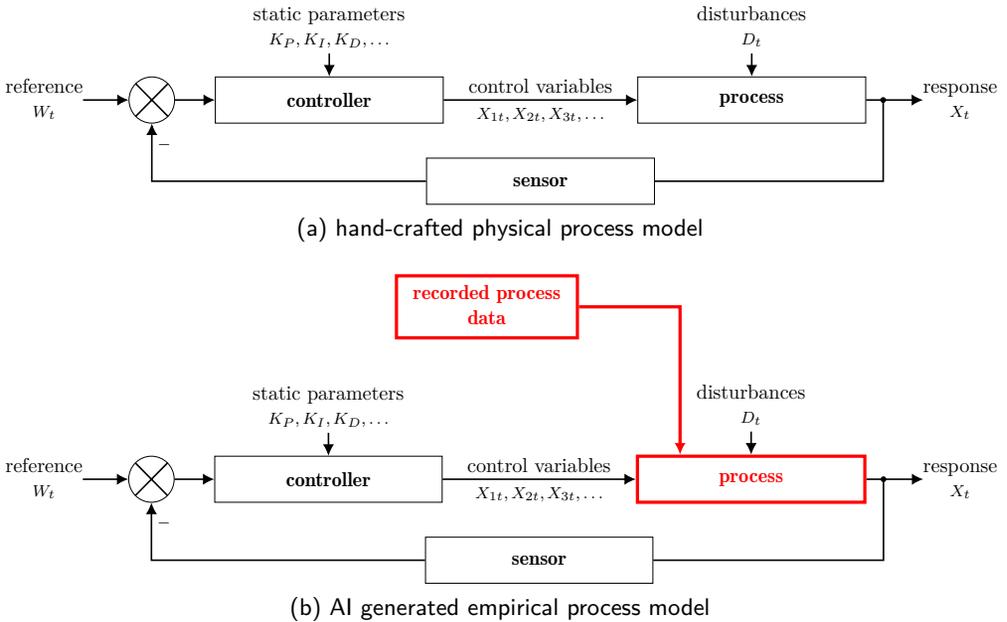

\subsection{AI-based Process Modeling}\label{sec:ProcessModeling}
By starting any control system design, the physical model of the process is studied, modeled, scaled, and evaluated. These four steps are repeated until the design model fits the real-world process, which needs to be controlled by the control system. Regarding to the statement by Hornik et al. \cite{Hornik1989}, that "standard multilayer feedforward networks are capable of approximating any measurable function to any desired degree of accuracy, in a very specific and satisfying sense", it seems to be obvious to replace the physical model with an empirical model. As illustrated in Figure \ref{fig:CLS_pyh_emp} \subref{fig:CLS_emp}, an ANN is trained on recorded data of the process. Thus, the trained ANN as functions approximator is an empirical model of the process model. Its level of detail depends on the training data, but the needs during the controller design are met with sufficient training data. Note, in control system engineering, all data and datasets are time series; thus, sampling frequencies during data recording need to be treated during learning.

The level of detail of the AI-generated empirical process model corresponds with the amount, the variance, and the biases of the recorded training data. This means, in case the recorded data cover every assumed process variance without any biases, the realistic process model can be automated generated.

Since it is challenging to collect data from the process in border and extreme cases, a mixture of both an empirical and a physical process model is needed. Physics-informed neural networks \cite{Raissi2019} as well as neural operator \cite{Li2020} and their combinations \cite{Li2021} are promising approaches for combining these both models. Here, a physical model definition of the process covers the border and extreme cases, while the training data is considered for the general cases during the model training.

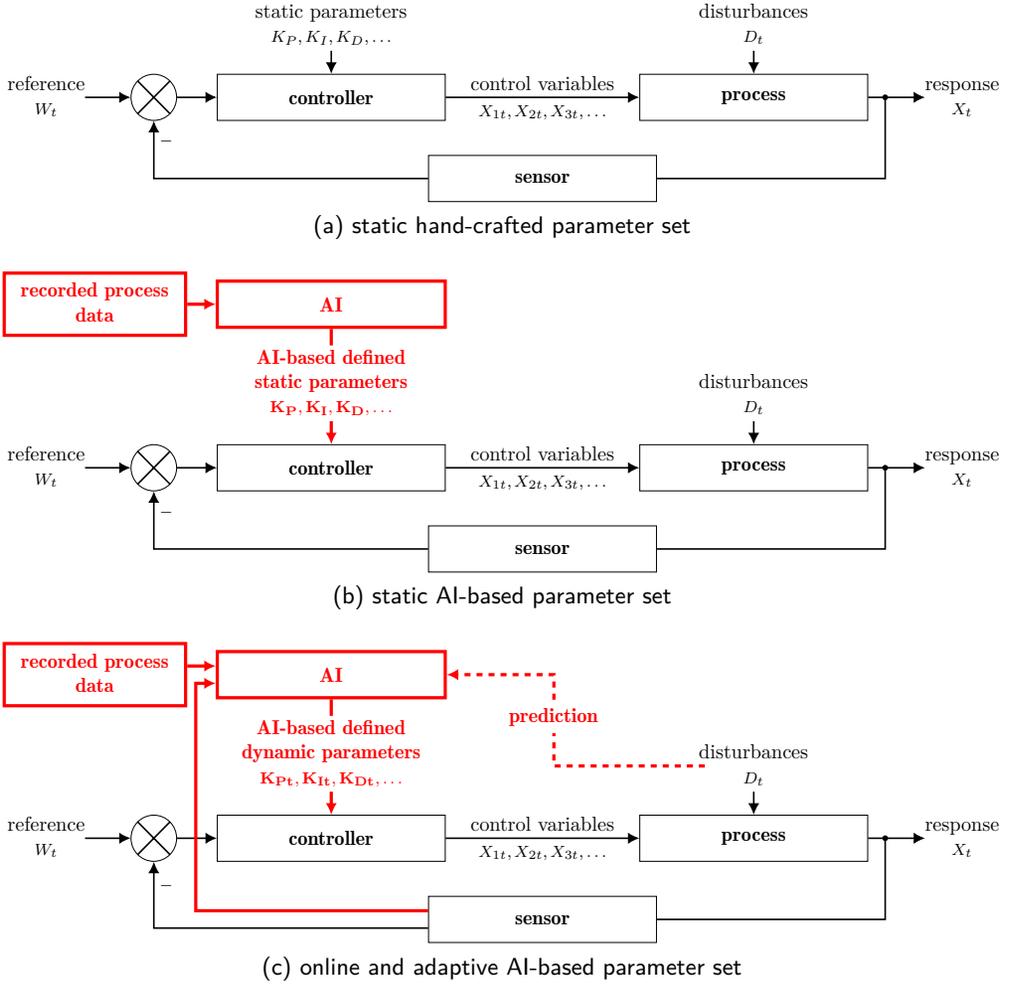
\begin{figure}[htb!]
	\centering
	\subfloat[static hand-crafted parameter set  \label{fig:CLS_a}] {
  \resizebox{\linewidth}{!}
  {
  \begin{tikzpicture}
  \node(controller)[draw, rectangle, minimum width = 4.0cm, minimum height = 0.8cm] {\textbf{controller}};

  \node(parameter)[above = 0.3cm of controller, rectangle, minimum height = 0.8cm] {\begin{tabular}{c}
  static parameters \\ \footnotesize $(K_{P},K_{I},K_{D},\ldots)$
  \end{tabular}
  };

  \node(process)[draw, right = 3.4cm of controller, rectangle, minimum width = 4.0cm, minimum height = 0.8cm] {\textbf{process}};

  \node(disturbances)[above = 0.3cm of process, rectangle, minimum height = 0.8cm] {\begin{tabular}{c}
  disturbances \\ \footnotesize $(D_{(t)})$
  \end{tabular}
  };

  \node(sensor)[draw, below right = 1cm and 1.7cm of controller, rectangle, minimum width = 4.0cm, minimum height = 0.8cm, anchor=center] {\textbf{sensor}};

  \node(response)[right = 0.7cm of process, rectangle, minimum height = 0.8cm] {\begin{tabular}{c}
   response \\ \footnotesize $(X_{(t)})$
  \end{tabular}
  };
  \node(sum)[draw, left = 0.7cm of controller, circle, minimum height = 0.8cm] {
  };
  \draw [thick] (sum.north east) -- (sum.south west)
  (sum.north west) -- (sum.south east);

  \node(reference)[left = 0.5cm of sum, rectangle, minimum height = 0.8cm] {\begin{tabular}{c}
   reference \\ \footnotesize $(W_{(t)})$
  \end{tabular}
  };

  \draw [-{Latex[length=2mm, width=2mm]}, thick, draw=black](controller.east)--(process.west) node [above,pos=0.5] (controlVariables) {control variables} node [below,pos=0.5] {\footnotesize $(X_{1(t)},X_{2(t)},X_{3(t)},\ldots)$};
  \draw [-{Latex[length=2mm, width=2mm]}, thick, draw=black](process.east)--($(response.west)+(0.3,0)$) node (rs) [circle, pos=0.3, fill=black, inner sep=0pt,minimum size=1mm] {};
  \draw [-{Latex[length=2mm, width=2mm]}, thick, draw=black](sum.east)--(controller.west);
  \draw [-{Latex[length=2mm, width=2mm]}, thick, draw=black]($(reference.east)+(-0.3,0)$)--(sum.west);
  \draw [-{Latex[length=2mm, width=2mm]}, thick, draw=black]($(disturbances.south)+(0,0.1)$)--(process.north);
  \draw [-{Latex[length=2mm, width=2mm]}, thick, draw=black]($(parameter.south)+(0,0.1)$)--(controller.north);
  \draw [thick, draw=black](rs.center)|-(sensor.east);
  \draw [-{Latex[length=2mm, width=2mm]}, thick, draw=black](sensor.west)-|(sum.south) node [right,pos=0.8] {\textbf{--}};

  \end{tikzpicture}
  }
  }

  \subfloat[static AI-based parameter set \label{fig:CLS_b}] {
  \resizebox{\linewidth}{!}
  {
  \begin{tikzpicture}
  \node(controller)[draw, rectangle, minimum width = 4.0cm, minimum height = 0.8cm] {\textbf{controller}};

  \node(parameter)[above = 0.3cm of controller, rectangle, minimum height = 0.8cm] {\color{red}{\begin{tabular}{c}
  \textbf{AI-based defined} \\ \textbf{static parameters} \\ \footnotesize $\mathbf{(K_{P},K_{I},K_{D},\ldots)}$
  \end{tabular}}
  };

  \node(process)[draw, right = 3.4cm of controller, rectangle, minimum width = 4.0cm, minimum height = 0.8cm] {\textbf{process}};

  \node(disturbances)[above = 0.3cm of process, rectangle, minimum height = 0.8cm] {\begin{tabular}{c}
  disturbances \\ \footnotesize $(D_{(t)})$
  \end{tabular}
  };

  \node(sensor)[draw, below right = 1cm and 1.7cm of controller, rectangle, minimum width = 4.0cm, minimum height = 0.8cm, anchor=center] {\textbf{sensor}};

  \node(response)[right = 0.7cm of process, rectangle, minimum height = 0.8cm] {\begin{tabular}{c}
   response \\ \footnotesize $(X_{(t)})$
  \end{tabular}
  };
  \node(sum)[draw, left = 0.7cm of controller, circle, minimum height = 0.8cm] {
  };
  \draw [thick] (sum.north east) -- (sum.south west)
  (sum.north west) -- (sum.south east);

  \node(reference)[left = 0.5cm of sum, rectangle, minimum height = 0.8cm] {\begin{tabular}{c}
   reference \\ \footnotesize $(W_{(t)})$
  \end{tabular}
  };

  \draw [-{Latex[length=2mm, width=2mm]}, thick, draw=black](controller.east)--(process.west) node [above,pos=0.5] (controlVariables) {control variables} node [below,pos=0.5] {\footnotesize $(X_{1(t)},X_{2(t)},X_{3(t)},\ldots)$};
  \draw [-{Latex[length=2mm, width=2mm]}, thick, draw=black](process.east)--($(response.west)+(0.3,0)$) node (rs) [circle, pos=0.3, fill=black, inner sep=0pt,minimum size=1mm] {};
  \draw [-{Latex[length=2mm, width=2mm]}, thick, draw=black](sum.east)--(controller.west);
  \draw [-{Latex[length=2mm, width=2mm]}, thick, draw=black]($(reference.east)+(-0.3,0)$)--(sum.west);
  \draw [-{Latex[length=2mm, width=2mm]}, thick, draw=black]($(disturbances.south)+(0,0.1)$)--(process.north);
  \draw [-{Latex[length=2mm, width=2mm]}, ultra thick, draw=red]($(parameter.south)+(0,0.1)$)--(controller.north);
  \draw [thick, draw=black](rs.center)|-(sensor.east);
  \draw [-{Latex[length=2mm, width=2mm]}, thick, draw=black](sensor.west)-|(sum.south) node [right,pos=0.8] {\textbf{--}};

  \node(AI)[draw=red, ultra thick, above = 2cm of controller, rectangle, minimum width = 4.0cm, minimum height = 0.8cm] {\color{red}{\textbf{AI}}};

  \node(recordedData)[draw=red, ultra thick, left = 0.5cm of AI, rectangle, minimum width = 3.0cm, minimum height = 0.8cm] {\color{red}{\begin{tabular}{c}\textbf{recorded process}\\ \textbf{data}
  \end{tabular}
  }};

  \draw [-{Latex[length=2mm, width=2mm]}, draw=red, ultra thick] (recordedData.east)--(AI.west);

  \draw [draw=red, ultra thick]($(parameter.north)-(0,0.1)$)--(AI.south);

  \end{tikzpicture}
  }
  }

  \subfloat[online and adaptive AI-based parameter set \label{fig:CLS_c}] {
  \resizebox{\linewidth}{!}
  {
  \begin{tikzpicture}
  \node(controller)[draw, rectangle, minimum width = 4.0cm, minimum height = 0.8cm] {\textbf{controller}};

  \node(parameter)[above = 0.3cm of controller, rectangle, minimum height = 0.8cm] {\color{red}{\begin{tabular}{c}
  \textbf{AI-based defined} \\ \textbf{dynamic parameters} \\ \footnotesize $\mathbf{(K_{P(t)},K_{I(t)},K_{D(t)},\ldots)}$
  \end{tabular}}
  };

  \node(process)[draw, right = 3.4cm of controller, rectangle, minimum width = 4.0cm, minimum height = 0.8cm] {\textbf{process}};

  \node(disturbances)[above = 0.3cm of process, rectangle, minimum height = 0.8cm] {\begin{tabular}{c}
  disturbances \\ \footnotesize $(D_{(t)})$
  \end{tabular}
  };

  \node(sensor)[draw, below right = 1cm and 1.7cm of controller, rectangle, minimum width = 4.0cm, minimum height = 0.8cm, anchor=center] {\textbf{sensor}};

  \node(response)[right = 0.7cm of process, rectangle, minimum height = 0.8cm] {\begin{tabular}{c}
   response \\ \footnotesize $(X_{(t)})$
  \end{tabular}
  };
  \node(sum)[draw, left = 0.7cm of controller, circle, minimum height = 0.8cm] {
  };
  \draw [thick] (sum.north east) -- (sum.south west)
  (sum.north west) -- (sum.south east);

  \node(reference)[left = 0.5cm of sum, rectangle, minimum height = 0.8cm] {\begin{tabular}{c}
   reference \\ \footnotesize $(W_{(t)})$
  \end{tabular}
  };

  \draw [-{Latex[length=2mm, width=2mm]}, thick, draw=black](controller.east)--(process.west) node [above,pos=0.5] (controlVariables) {control variables} node [below,pos=0.5] {\footnotesize $(X_{1(t)},X_{2(t)},X_{3(t)},\ldots)$};
  \draw [-{Latex[length=2mm, width=2mm]}, thick, draw=black](process.east)--($(response.west)+(0.3,0)$) node (rs) [circle, pos=0.3, fill=black, inner sep=0pt,minimum size=1mm] {};
  \draw [-{Latex[length=2mm, width=2mm]}, thick, draw=black](sum.east)--(controller.west);
  \draw [-{Latex[length=2mm, width=2mm]}, thick, draw=black]($(reference.east)+(-0.3,0)$)--(sum.west);
  \draw [-{Latex[length=2mm, width=2mm]}, thick, draw=black]($(disturbances.south)+(0,0.1)$)--(process.north);
  \draw [-{Latex[length=2mm, width=2mm]}, ultra thick, draw=red]($(parameter.south)+(0,0.1)$)--(controller.north);
  \draw [thick, draw=black](rs.center)|-(sensor.east);
  \draw [-{Latex[length=2mm, width=2mm]}, thick, draw=black]($(sensor.west)-(0,0.15)$)-|(sum.south) node [right,pos=0.8] {\textbf{--}};

  \node(AI)[draw=red, ultra thick, above = 2cm of controller, rectangle, minimum width = 4.0cm, minimum height = 0.8cm] {\color{red}{\textbf{AI}}};

  \node(recordedData)[draw=red, ultra thick, left = 0.5cm of AI, rectangle, minimum width = 3.0cm, minimum height = 0.8cm] {\color{red}{\begin{tabular}{c}\textbf{recorded process}\\ \textbf{data}
  \end{tabular}
  }};

  \draw [-{Latex[length=2mm, width=2mm]}, draw=red, ultra thick] ($(recordedData.east)+(0,0.15)$)--($(AI.west)+(0,0.15)$);

  \coordinate [above left = 1.7cm and 1.5cm of process]  (predictionText);

  \draw [-{Latex[length=2mm, width=2mm]}, draw=red, ultra thick, dashed] ($(disturbances.west)+(0.4,0)$)-|(predictionText)|-(AI.east);

  \node [fill=white] (prediction) at (predictionText){\color{red}{\textbf{prediction}}};

  \coordinate [below left = 0.75cm and 0.35cm of AI]  (sensorFeedbackAI);

  \draw [-{Latex[length=2mm, width=2mm]}, draw=red, ultra thick] ($(sensor.west)+(0,0.15)$)-|(sensorFeedbackAI)|-($(AI.west)-(0,0.15)$);

  \draw [draw=red, ultra thick]($(parameter.north)-(0,0.1)$)--(AI.south);

  \end{tikzpicture}
  }
  }
\caption{Parameter tuning for CLCS,  \protect\subref{fig:CLS_a}  fine tuned parameters by experts,  \protect\subref{fig:CLS_b} and  \protect\subref{fig:CLS_c} process data driven tuned adaptive parameters.
}
\label{fig:CLS}
\end{figure}

\subsection{AI-empowered and -based Parameter Tuning}
To define the optional parameter set, e.g., for PID controllers, Ziegler-Nichols tuning, Cohen-Coon tuning, Kappa-Tau tuning, heuristic tuning, and other methods are used nowadays. These methodologies lead to parameter sets, which are then commonly fine-tuned by hand, resulting in static hand-crafted control parameter, as seen in Figure \ref{fig:CLS} \subref{fig:CLS_a}.

Next to the established tuning approaches, ANN can optimize the parameter for typical process cases based on recorded data. Illustrated in Figure \ref{fig:CLS} \subref{fig:CLS_b}, the parameter for the controller are trained as static parameter sets so that these AI-tuned parameters can be evaluated before deploying the CLCS, ensuring that the parameter set coves the extreme cases. The advantage of AI-based, i.e., data-driven parameter tuning, is that the probability distribution of process states and the probability distribution of disturbance will be considered during training. Because, unlike the dataset for the AI-based process modeling, the data set used for parameter tuning should represent the real-world process cases. Thus the data set for parameter tuning might be strongly biased with recurrent process states.

Based on the recorded process data, an ANN can be trained for proving a dynamic set of parameter to the controller, visualized in Figure \ref{fig:CLS} \subref{fig:CLS_c}. In this setting, the ANN might internally predict the disturbance for generating the best parameter set online. Increasing this internal representation of the current process state and the current disturbance within the ANN, the ANN needs to get sensor input directly. This sensor input can originate from the sensor, used for the feedback of the CLCS, but other additional sensors can be considered as well. With multiple sensors, the prediction of process disturbances can be optimized by training the ANN. Thus, the ANN can dynamically estimate the optimal parameter set and update in the controller. Online parameter tuning in non-convex optimization cases becomes for particular controllers, like the PID controller, mathematically an NP-hard problem \cite{Somefun2021}. Introducing AI, in the form of ANN might deal with the issue of NP-hard calculations \cite{Memon2020} during runtime.

In contrast to expert rule tables for dynamical parameter tuning \cite{Liu2020}, the online estimation of controller parameters by an ANN leads to an enormous space of possible CLCS characteristics. All possible system behaviors resulting from the enormous working space of CLCS with online AI-based parameter tuning can not be simulated or tested in entirety. Consequently, AI-based dynamically online change of controller parameter sets is not recommended yet for time- and safety-critical processes.

\begin{figure}[htb!]
	\centering
  \subfloat[AI-based controller \label{fig:CLS_ai}] {
  \resizebox{\linewidth}{!}
  {
  \begin{tikzpicture}
  \node(controller)[draw=red, ultra thick, rectangle, minimum width = 4.0cm, minimum height = 0.8cm] {\color{red}{\textbf{AI}}};


  \node(process)[draw, right = 3.4cm of controller, rectangle, minimum width = 4.0cm, minimum height = 0.8cm] {\textbf{process}};

  \node(disturbances)[above = 0.3cm of process, rectangle, minimum height = 0.8cm] {\begin{tabular}{c}
  disturbances \\ \footnotesize $(D_{(t)})$
  \end{tabular}
  };

  \node(sensor)[draw, below right = 1cm and 1.7cm of controller, rectangle, minimum width = 4.0cm, minimum height = 0.8cm, anchor=center] {\textbf{sensor}};

  \node(response)[right = 0.7cm of process, rectangle, minimum height = 0.8cm] {\begin{tabular}{c}
   response \\ \footnotesize $(X_{(t)})$
  \end{tabular}
  };
  \node(sum)[left = 0.7cm of controller, circle, minimum height = 0.8cm] {
  };

  \node(reference)[left = 0.5cm of sum, rectangle, minimum height = 0.8cm] {\begin{tabular}{c}
   reference \\ \footnotesize $(W_{(t)})$
  \end{tabular}
  };

  \draw [-{Latex[length=2mm, width=2mm]}, thick, draw=black](controller.east)--(process.west) node [above,pos=0.5] (controlVariables) {control variables} node [below,pos=0.5] {\footnotesize $(X_{1(t)},X_{2(t)},X_{3(t)},\ldots)$};
  \draw [-{Latex[length=2mm, width=2mm]}, thick, draw=black](process.east)--($(response.west)+(0.3,0)$) node (rs) [circle, pos=0.3, fill=black, inner sep=0pt,minimum size=1mm] {};
  \draw [-{Latex[length=2mm, width=2mm]}, thick, draw=black]($(reference.east)+(-0.3,0)$)--(controller.west);
  \draw [-{Latex[length=2mm, width=2mm]}, thick, draw=black]($(disturbances.south)+(0,0.1)$)--(process.north);
  \draw [thick, draw=black](rs.center)|-(sensor.east);
  \draw [-{Latex[length=2mm, width=2mm]}, ultra thick, draw=red](sensor.west)-|(controller.south);

  \node(AI)[above = 2cm of controller, rectangle, minimum width = 4.0cm, minimum height = 0.8cm] {};

  \node(recordedData)[draw=red, ultra thick, left = 0.5cm of AI, rectangle, minimum width = 3.0cm, minimum height = 0.8cm] {\color{red}{\begin{tabular}{c}\textbf{recorded process}\\ \textbf{data}
  \end{tabular}
  }};

  \draw [-{Latex[length=2mm, width=2mm]}, draw=red, ultra thick] (recordedData.east)-|($(controller.north)-(0.15,0)$);

  \draw [-{Latex[length=2mm, width=2mm]}, draw=red, ultra thick, dashed] ($(disturbances.west)+(0.4,0)$)-|($(controller.north)+(0.15,0)$)node [above, pos=0.25] {\color{red}{\textbf{prediction}}};

  \end{tikzpicture}
  }
  }

	\subfloat[cascaded setup with static parameter sets \label{fig:CLS_ml_a}] {
  \resizebox{\linewidth}{!}
  {
  \begin{tikzpicture}
  \node(controller)[draw, rectangle, minimum width = 2.35cm, minimum height = 0.8cm] {\textbf{controller}};

  \node(parameter)[above = 0.3cm of controller, rectangle, minimum height = 0.8cm] {\begin{tabular}{c}
  static parameters \\ \footnotesize $(K_{P},K_{I},K_{D},\ldots)$
  \end{tabular}
  };

  \node(sum2)[draw, right = 0.4cm of controller, circle, minimum height = 0.8cm] {
  };
  \draw [thick] (sum2.north east) -- (sum2.south west)
  (sum2.north west) -- (sum2.south east);

  \node(controller2)[draw, right = 0.4cm of sum2, rectangle, minimum width = 2.35cm, minimum height = 0.8cm] {\textbf{controller}};

  \node(parameter2)[above = 0.3cm of controller2, rectangle, minimum height = 0.8cm] {\begin{tabular}{c}
  static parameters \\ \footnotesize $(K_{P},K_{I},K_{D},\ldots)$
  \end{tabular}
  };

  \node(process)[draw, right = 3.4cm of controller2, rectangle, minimum width = 2.0cm, minimum height = 0.8cm] {\textbf{process}};

  \node(disturbances)[above = 0.3cm of process, rectangle, minimum height = 0.8cm] {\begin{tabular}{c}
  disturbances \\ \footnotesize $(D_{(t)})$
  \end{tabular}
  };

  \node(sensor)[draw, below right = 1cm and 1.7cm of controller2, rectangle, minimum width = 4cm, minimum height = 0.8cm, anchor=center] {\textbf{sensor A}};

  \node(sensor2)[draw, below = 0.5cm of sensor, rectangle, minimum width = 4cm, minimum height = 0.8cm, anchor=center] {\textbf{sensor B}};

  \node(response)[right = 0.7cm of process, rectangle, minimum height = 0.8cm] {\begin{tabular}{c}
   response \\ \footnotesize $(X_{(t)})$
  \end{tabular}
  };
  \node(sum)[draw, left = 0.4cm of controller, circle, minimum height = 0.8cm] {
  };
  \draw [thick] (sum.north east) -- (sum.south west)
  (sum.north west) -- (sum.south east);

  \node(reference)[left = 0.5cm of sum, rectangle, minimum height = 0.8cm] {\begin{tabular}{c}
   reference \\ \footnotesize $(W_{(t)})$
  \end{tabular}
  };

  \draw [-{Latex[length=2mm, width=2mm]}, thick, draw=black](controller2.east)--(process.west) node [above,pos=0.5] (controlVariables) {control variables} node [below,pos=0.5] {\footnotesize $(X_{1(t)},X_{2(t)},X_{3(t)},\ldots)$};
  \draw [-{Latex[length=2mm, width=2mm]}, thick, draw=black](process.east)--($(response.west)+(0.3,0)$) node (rs) [circle, pos=0.3, fill=black, inner sep=0pt,minimum size=1mm] {};
  \draw [-{Latex[length=2mm, width=2mm]}, thick, draw=black](sum.east)--(controller.west);
  \draw [-{Latex[length=2mm, width=2mm]}, thick, draw=black](controller.east)--(sum2.west);
  \draw [-{Latex[length=2mm, width=2mm]}, thick, draw=black](sum2.east)--(controller2.west);
  \draw [-{Latex[length=2mm, width=2mm]}, thick, draw=black]($(reference.east)+(-0.3,0)$)--(sum.west);
  \draw [-{Latex[length=2mm, width=2mm]}, thick, draw=black]($(disturbances.south)+(0,0.1)$)--(process.north);
  \draw [-{Latex[length=2mm, width=2mm]}, thick, draw=black]($(parameter.south)+(0,0.1)$)--(controller.north);
  \draw [-{Latex[length=2mm, width=2mm]}, thick, draw=black]($(parameter2.south)+(0,0.1)$)--(controller2.north);
  \draw [thick, draw=black](rs.center)|-(sensor.east);
  \draw [thick, draw=black](rs.center)|-(sensor2.east);
  \draw [-{Latex[length=2mm, width=2mm]}, thick, draw=black](sensor.west)-|(sum2.south) node [right,pos=0.8] {\textbf{--}};
  \draw [-{Latex[length=2mm, width=2mm]}, thick, draw=black](sensor2.west)-|(sum.south) node [right,pos=0.85] {\textbf{--}};

  \end{tikzpicture}
  }
  }

  \subfloat[simplification due to AI-based controller \label{fig:CLS_ml_b}] {
  \resizebox{\linewidth}{!}
  {
  \begin{tikzpicture}
  \node(controller)[draw=red, ultra thick, rectangle, minimum width = 4.0cm, minimum height = 0.8cm] {\color{red}{\textbf{AI}}};


  \node(process)[draw, right = 3.4cm of controller, rectangle, minimum width = 4.0cm, minimum height = 0.8cm] {\textbf{process}};

  \node(disturbances)[above = 0.3cm of process, rectangle, minimum height = 0.8cm] {\begin{tabular}{c}
  disturbances \\ \footnotesize $(D_{(t)})$
  \end{tabular}
  };

  \node(sensor)[draw, below right = 1cm and 1.7cm of controller, rectangle, minimum width = 4cm, minimum height = 0.8cm, anchor=center] {\textbf{sensor A}};

  \node(sensor2)[draw, below = 0.5cm of sensor, rectangle, minimum width = 4cm, minimum height = 0.8cm, anchor=center] {\textbf{sensor B}};

  \node(response)[right = 0.7cm of process, rectangle, minimum height = 0.8cm] {\begin{tabular}{c}
   response \\ \footnotesize $(X_{(t)})$
  \end{tabular}
  };
  \node(sum)[left = 0.7cm of controller, circle, minimum height = 0.8cm] {
  };

  \node(reference)[left = 0.5cm of sum, rectangle, minimum height = 0.8cm] {\begin{tabular}{c}
   reference \\ \footnotesize $(W_{(t)})$
  \end{tabular}
  };

  \draw [-{Latex[length=2mm, width=2mm]}, thick, draw=black](controller.east)--(process.west) node [above,pos=0.5] (controlVariables) {control variables} node [below,pos=0.5] {\footnotesize $(X_{1(t)},X_{2(t)},X_{3(t)},\ldots)$};
  \draw [-{Latex[length=2mm, width=2mm]}, thick, draw=black](process.east)--($(response.west)+(0.3,0)$) node (rs) [circle, pos=0.3, fill=black, inner sep=0pt,minimum size=1mm] {};
  \draw [-{Latex[length=2mm, width=2mm]}, thick, draw=black]($(reference.east)+(-0.3,0)$)--(controller.west);
  \draw [-{Latex[length=2mm, width=2mm]}, thick, draw=black]($(disturbances.south)+(0,0.1)$)--(process.north);
  \draw [thick, draw=black](rs.center)|-(sensor.east);
  \draw [thick, draw=black](rs.center)|-(sensor2.east);
  \draw [-{Latex[length=2mm, width=2mm]}, ultra thick, draw=red](sensor.west)-|($(controller.south)+(0.15,0)$);
  \draw [-{Latex[length=2mm, width=2mm]}, ultra thick, draw=red](sensor2.west)-|($(controller.south)-(0.15,0)$);

  \node(AI)[above = 2cm of controller, rectangle, minimum width = 4.0cm, minimum height = 0.8cm] {};

  \node(recordedData)[draw=red, ultra thick, left = 0.5cm of AI, rectangle, minimum width = 3.0cm, minimum height = 0.8cm] {\color{red}{\begin{tabular}{c}\textbf{recorded process}\\ \textbf{data}
  \end{tabular}
  }};

  \draw [-{Latex[length=2mm, width=2mm]}, draw=red, ultra thick] (recordedData.east)-|($(controller.north)-(0.15,0)$);

  \draw [-{Latex[length=2mm, width=2mm]}, draw=red, ultra thick, dashed] ($(disturbances.west)+(0.4,0)$)-|($(controller.north)+(0.15,0)$)node [above, pos=0.25] {\color{red}{\textbf{prediction}}};

  \end{tikzpicture}
  }
  }

\caption{AI-based controller: \protect\subref{fig:CLS_ai}, e.g.,  an ANN replaces the controller design; \protect\subref{fig:CLS_ml_a} and \protect\subref{fig:CLS_ml_b} cascaded controller set up where an AI-bases controller like in \protect\subref{fig:CLS_ml_b} simplifies the CLCS design.}
\label{fig:CLS_ml}
\end{figure}
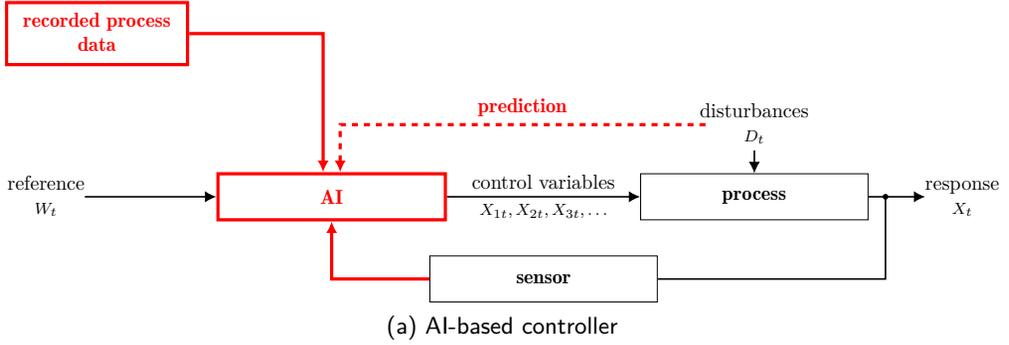
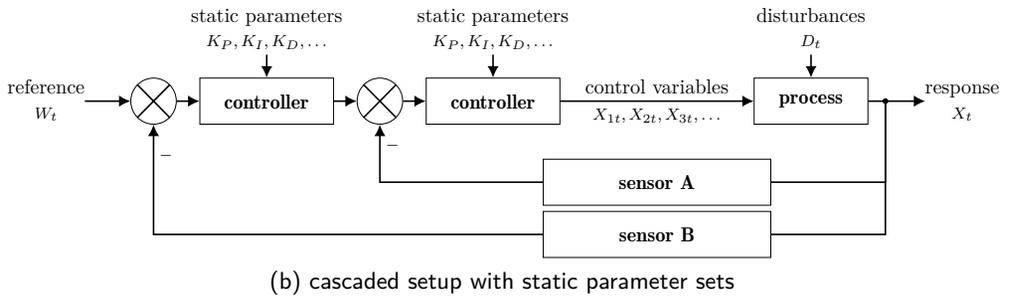
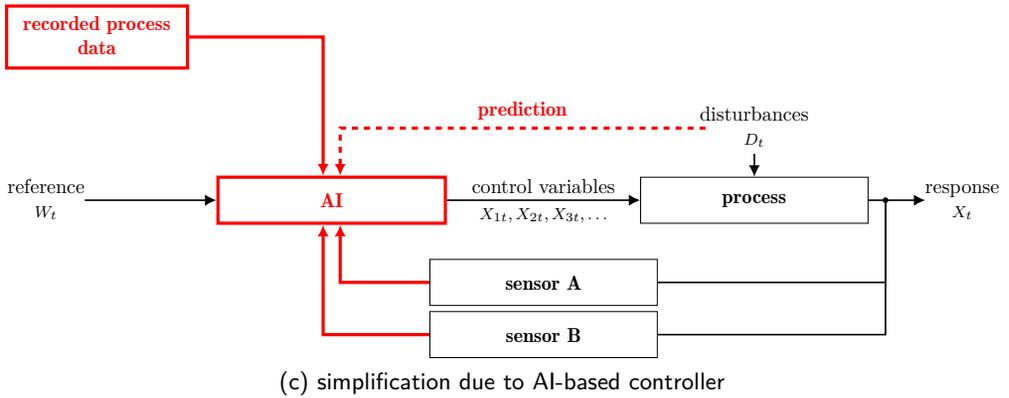

\subsection{AI-based Controller}
Going even a step further, the entire controller, i.e., even nested \cite{Voulgaris2000}, and cascaded \cite{Lee1998,Jeng2012} controller setting, can be replaced by AI-based controller. As shown in the block diagram of Figure \ref{fig:CLS_ml} \subref{fig:CLS_ai}, a trained AI-based controller, most likely in the form of an ANN-based controller, directly sets the control variables based on the given reference value and the sensor input. Even multiple reference values, as well as sensor inputs, are directly coupled to the AI-based controller. The challenging part of CLCS design with an AI-based controller is now the training of AI. Therefore, real-world data with the typical process variance are needed and an unbiased data set, covering all variance without bias. Conserving the characteristic of both time-series data sets during learning is challenging; a possible solution is to train on both datasets simultaneously \cite{Gu2020,Ye2021}.

In contrast, to the complex and time-expensive training of an AI-based controller, the inference times, i.e., the execution times, of trained ANN are magnitudes faster. On GPU and particular FPGA, the inference times of deep ANN are below $10ms$ \cite{Duarte2018,KerasTeam2021} and fit many processes in control. The inference times of lightweight ANNs with less than four million parameter are usual below $3.9ms$ \cite{KerasTeam2021}. AI-based controllers can no longer be evaluated entirely from an evaluation and testing perspective. Even by systemically evaluating the system behavior in predefined input-values spaces, poles and extra can remain unnoticed.

\begin{figure}[htb!]
	\centering
  \resizebox{\linewidth}{!}
  {
  \begin{tikzpicture}
  \node(controller)[draw=red, ultra thick, rectangle, minimum width = 4.0cm, minimum height = 0.8cm] {\color{red}{\textbf{AI}}};


  \node(process)[draw=red, ultra thick, right = 3.4cm of controller, rectangle, minimum width = 4.0cm, minimum height = 0.8cm] {\color{red}{\textbf{process}}};

  \node(disturbances)[above = 0.3cm of process, rectangle, minimum height = 0.8cm] {\begin{tabular}{c}
  disturbances \\ \footnotesize $(D_{(t)})$
  \end{tabular}
  };

  \node(sensor)[draw, below right = 1cm and 1.7cm of controller, rectangle, minimum width = 4.0cm, minimum height = 0.8cm, anchor=center] {\textbf{sensor}};

  \node(response)[right = 0.7cm of process, rectangle, minimum height = 0.8cm] {\begin{tabular}{c}
   response \\ \footnotesize $(X_{(t)})$
  \end{tabular}
  };
  \node(sum)[left = 0.7cm of controller, circle, minimum height = 0.8cm] {
  };

  \node(reference)[left = 0.5cm of sum, rectangle, minimum height = 0.8cm] {\begin{tabular}{c}
   reference \\ \footnotesize $(W_{(t)})$
  \end{tabular}
  };

  \draw [-{Latex[length=2mm, width=2mm]}, thick, draw=black](controller.east)--(process.west) node [above,pos=0.5] (controlVariables) {control variables} node [below,pos=0.5] {\footnotesize $(X_{1(t)},X_{2(t)},X_{3(t)},\ldots)$};
  \draw [-{Latex[length=2mm, width=2mm]}, thick, draw=black](process.east)--($(response.west)+(0.3,0)$) node (rs) [circle, pos=0.3, fill=black, inner sep=0pt,minimum size=1mm] {};
  \draw [-{Latex[length=2mm, width=2mm]}, thick, draw=black]($(reference.east)+(-0.3,0)$)--(controller.west);
  \draw [-{Latex[length=2mm, width=2mm]}, thick, draw=black]($(disturbances.south)+(0,0.1)$)--(process.north);
  \draw [thick, draw=black](rs.center)|-(sensor.east);
  \draw [-{Latex[length=2mm, width=2mm]}, ultra thick, draw=red](sensor.west)-|(controller.south);

  \node(processText)[above = 1.5cm of process, rectangle] {\color{red}\footnotesize{\begin{minipage}{6.5cm}\begin{itemize}
      \item[\textit{I)}] record data set of process and create process model, empirically, by AI
\end{itemize}\end{minipage}}
  };
  \draw[decorate,decoration={brace,amplitude=5pt, aspect=0.25}, draw=red] ($(processText.south east)+(0,0.1)$) -- node[below=0pt, pos=0.25] (processCBrace) {}  ($(processText.south west)+(0,0.1)$);

  \draw [-{Latex[length=2mm, width=2mm]}, dashed, draw=red] let
  \p1=($(process.north)$), \p2=($(processCBrace.center)$) in (processCBrace.center)--(\x2,\y1);

  \node(controlText)[above = 0.2cm of controlVariables, rectangle] {\color{blue}\footnotesize{\begin{minipage}{3.9cm}\begin{itemize}
      \item[\textit{II)}] decide which variables are to be controlled
      \item[\textit{III)}] decide on the actuators
\end{itemize}\end{minipage}}
  };
  \draw[decorate,decoration={brace,amplitude=5pt, aspect=0.14}, draw=blue] ($(controlText.south east)+(0,0.1)$) -- node[below=0pt, pos=0.14] (controlCBrace) {}  ($(controlText.south west)+(0,0.1)$);

  \draw [-{Latex[length=2mm, width=2mm]}, dashed, draw=blue] let
  \p1=($(process.east)$), \p2=($(controlCBrace.center)$) in (controlCBrace.center)--(\x2,\y1);

  \node(measurementsText)[below right = 1.85cm and -2.225cm of process, rectangle] {\color{blue}\footnotesize{\begin{minipage}{4.5cm}\begin{itemize}
      \item[\textit{III)}] decide on the measurements
\end{itemize}\end{minipage}}
  };
  \draw[decorate,decoration={brace,amplitude=5pt, aspect=0.5}, draw=blue] ($(measurementsText.north west)-(0,0.1)$) -- node[above=0pt, pos=0.5] (measurementsCBrace) {}  ($(measurementsText.north east)-(0,0.1)$);

  \draw [-{Latex[length=2mm, width=2mm]}, dashed, draw=blue] let
  \p1=($(process.west)$), \p2=($(measurementsCBrace.center)$) in (measurementsCBrace.center)--(\x2,\y1);

  \node(sensorText)[left = 1.85cm of measurementsText, rectangle] {\color{blue}\footnotesize{\begin{minipage}{3.5cm}\begin{itemize}
      \item[\textit{III)}] decide on the sensors
  \end{itemize}\end{minipage}}
  };
  \draw[decorate,decoration={brace,amplitude=5pt, aspect=0.5}, draw=blue] ($(sensorText.north west)-(0,0.1)$) -- node[above=0pt, pos=0.5] (sensorCBrace) {}  ($(sensorText.north east)-(0,0.1)$);

  \draw [-{Latex[length=2mm, width=2mm]}, dashed, draw=blue] let
  \p1=($(sensor.south)$), \p2=($(sensorCBrace.center)$) in (sensorCBrace.center)--(\x2,\y1);

  \node(controllerText)[above = 1.5cm of controller, rectangle] {\color{red}\footnotesize{\begin{minipage}{5cm}\begin{itemize}
      \item[\textit{V)}] train AI on control objective

  \end{itemize}\end{minipage}}
  };
  \draw[decorate,decoration={brace,amplitude=5pt, aspect=0.5}, draw=red] ($(controllerText.south east)+(0,0.1)$) -- node[below=0pt, pos=0.5] (controlerCBrace) {}  ($(controllerText.south west)+(0,0.1)$);

  \draw [-{Latex[length=2mm, width=2mm]}, dashed, draw=red] let
  \p1=($(controller.north)$), \p2=($(controlerCBrace)$) in (controlerCBrace)--(\x2,\y1);

  \node(sumText)[above left = 0.5cm and -1.75cm of sum, rectangle] {\color{blue}\footnotesize{\begin{minipage}{4.8cm}\begin{itemize}
      \item[\textit{IV)}] 	decide on the control objectives
  \end{itemize}\end{minipage}}
  };
  \draw[decorate,decoration={brace,amplitude=5pt, aspect=0.32}, draw=blue] ($(sumText.south east)+(0,0.1)$) -- node[below=0pt, pos=0.32] (sumCBrace) {}  ($(sumText.south west)+(0,0.1)$);

  \draw [-{Latex[length=2mm, width=2mm]}, dashed, draw=blue] let
  \p1=($(sum.center)$), \p2=($(sumCBrace)$) in (sumCBrace)--(\x2,\y1);

  \draw[decorate,decoration={brace,amplitude=5pt, aspect=0.5}, draw=blue] let
  \p1=($(reference.west)$), \p2=($(response.east)$), \p3=($(sensorText.south)$) in ($(\x2,\y3)-(0,0.1)$) -- node[below=0pt, pos=0.5] (systemCBrace) {}  ($(\x1,\y3)-(0,0.1)$);

  \node(systemText)[below = 0.25cm of systemCBrace, rectangle] {\color{blue}\footnotesize{\begin{minipage}[t]{6.6cm}\begin{itemize}
      \item[\textit{V)}] simulate the controlled system
      \item[\textit{VI)}] repeat from step \textit{I)} if needed
      \item[\textit{VII)}] select the hardware for the controller
  \end{itemize}\end{minipage}
  \begin{minipage}[t]{6.6cm}\begin{itemize}
      \item[\textit{VIII)}] validate the controlled system in application
      \item[\textit{IX)}] repeat from step \textit{I)} or \textit{V)} if needed
  \end{itemize}\end{minipage} }
  };
  \draw[decorate,decoration={brace,amplitude=5pt, aspect=0.5}, draw=blue] ($(systemText.north west)-(0,0.1)$) -- node[above=0pt, pos=0.5] (systemIICBrace) {}  ($(systemText.north east)-(0,0.1)$);

  \draw [dashed, draw=blue] (systemCBrace)--(systemIICBrace);

  \end{tikzpicture}
  }
\caption{The process of AI-empowered and -based control system design, 9 steps in accordance with the corresponding block of the block diagram. Blue steps identical to the steps of Skogestad and Postlethwaite \cite{Skogestad2005} -- red steps are altered.}
\label{fig:CLS_process_AI}
\end{figure}
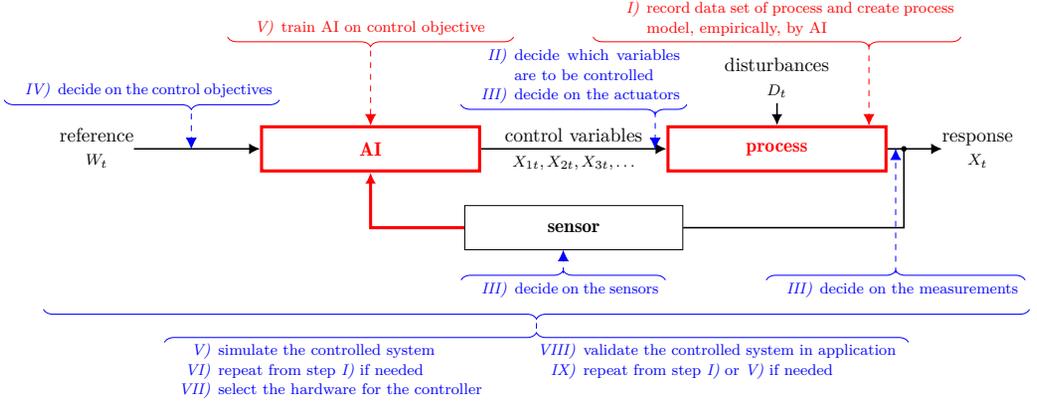

\subsection{Procedure of the AI-empowered and -based Control System Design}
Once AI can be used for and in CLCS design, the number of steps in its design will be reduced to nine steps.  As shown in Figure \ref{fig:CLS_process_AI}, the three steps needed for describing the physical process model are condensed to a single one---recording the data set and create the empirical model automatically by the use of AI. By using an AI controller, the decision and analysis on the controller type of as well as on the control configuration become obsolete. However, similar to the tuning of the controller, the training of the AI regarding the reference values is needed.

\section{Safety for AI-empowered and -based Closed-Loop Control Systems}\label{sec:safeCLCS}
AI and functional safety are nowadays mainly discussed in the context of obstacle detection for self-driving vehicles \cite{Zoeldy2020,Athavale2020,Tschurtz2021}, because high accurate obstacle detection on camera, LiDar, and radar data can only be achieved by AI algorithm. The safety dilemma of AI remains unsolved until explainable AI is detailed delved.

\begin{figure}[htb!]
	\centering
	\subfloat[AI-based controller with a conventional fall-back path \label{fig:CLS_safety_a}] {
  \resizebox{\linewidth}{!}
  {
  \begin{tikzpicture}
  \node(controller)[draw=red, ultra thick, rectangle, minimum width = 2.35cm, minimum height = 0.8cm] {\color{red}{\textbf{AI}}};

  \node(switch)[draw=red, ultra thick, right = 1.25cm of controller, rectangle, minimum width = 0.8cm, minimum height = 0.8cm] {};

  \node(AIswitch)[below = 0.3cm of switch, rectangle, minimum height = 0.8cm, inner sep=0pt]  {\color{red}\begin{tabular}{c}
  threshold \\ switch
  \end{tabular}
  };

  \node(process)[draw, right = 3.4cm of switch, rectangle, minimum width = 4.0cm, minimum height = 0.8cm] {\textbf{process}};

  \node(disturbances)[above = 0.3cm of process, rectangle, minimum height = 0.8cm] {\begin{tabular}{c}
  disturbances \\ \footnotesize $(D_{(t)})$
  \end{tabular}
  };

  \node(sensor)[draw, below right = 2.75cm and 3.35cm of controller, rectangle, minimum width = 4.0cm, minimum height = 0.8cm, anchor=center] {\textbf{sensor}};

  \node(response)[right = 0.7cm of process, rectangle, minimum height = 0.8cm] {\begin{tabular}{c}
   response \\ \footnotesize $(X_{(t)})$
  \end{tabular}
  };
  \node(sum)[left = 0.3cm of controller, circle, minimum height = 0.8cm] {
  };

  \node(reference)[left = 0.5cm of sum, rectangle, minimum height = 0.8cm] {\begin{tabular}{c}
   reference \\ \footnotesize $(W_{(t)})$
  \end{tabular}
  };

  \draw [-{Latex[length=2mm, width=2mm]}, ultra thick, draw=red]($(sensor.west)+(0,0.15)$)-|($(controller.south east)-(0.75,0)$);
  \node(controller2)[draw, below = 0.4cm of controller, thick, rectangle, minimum width = 2.35cm, minimum height = 0.8cm,fill=white] {{\textbf{controller}}};
  \node(parameter)[below = 0.3cm of controller2, rectangle, minimum height = 0.8cm, fill=white, inner sep=0pt]  {\begin{tabular}{c}
  static parameters \\ \footnotesize $(K_{P},K_{I},K_{D},\ldots)$
  \end{tabular}
  };

  \node(sum2)[draw, left = 0.3cm of controller2, circle, minimum height = 0.8cm] {
  };
  \draw [thick] (sum2.north east) -- (sum2.south west)
  (sum2.north west) -- (sum2.south east);

  \draw [-{Latex[length=2mm, width=2mm]}, thick, draw=black](switch.east)--(process.west) node [above,pos=0.5] (controlVariables) {control variables} node [below,pos=0.5] {\footnotesize $(X_{1(t)},X_{2(t)},X_{3(t)},\ldots)$};
  \draw [-{Latex[length=2mm, width=2mm]}, thick, draw=black](process.east)--($(response.west)+(0.3,0)$) node (rs) [circle, pos=0.3, fill=black, inner sep=0pt,minimum size=1mm] {};

  \draw [-{Latex[length=2mm, width=2mm]}, thick, draw=black]($(reference.east)+(-0.3,0)$)--(controller.west);
  \draw [-{Latex[length=2mm, width=2mm]}, thick, draw=black]($(disturbances.south)+(0,0.1)$)--(process.north);
  \draw [-{Latex[length=2mm, width=2mm]}, thick, draw=black]($(parameter.north)-(0,0.1)$)--(controller2.south);
  \draw [thick, draw=black](rs.center)|-(sensor.east);

  \draw [-{Latex[length=2mm, width=2mm]}, thick, draw=black]($(sensor.west)-(0,0.15)$)-|(sum2.south) node [right,pos=0.9] {\textbf{--}};

  \node (rsw) [left = 0.0cm of sum.center, circle, fill=black, inner sep=0pt,minimum size=1mm, anchor=center] {};
  \draw [-{Latex[length=2mm, width=2mm]}, ultra thick, draw=red](rsw.east)--(controller.west);
  \draw [-{Latex[length=2mm, width=2mm]},  thick, draw=black](rsw.center)--(sum2.north);
  \draw [-{Latex[length=2mm, width=2mm]}, thick, draw=black](sum2.east)--(controller2.west);

  \coordinate [left = 0.6cm of switch]  (swInput);

  \draw [-{Latex[length=2mm, width=2mm]}, ultra thick, draw=red](controller.east)-|($(swInput)+(0,0.15)$)|-($(switch.west)+(0,0.15)$);

  \draw [-{Latex[length=2mm, width=2mm]}, ultra thick, draw=red](controller2.east)-|($(swInput)-(0,0.15)$)|-($(switch.west)-(0,0.15)$);

  \coordinate [left = -0.2cm of switch]  (swIn);

  \node (swU) [above = 0.15cm of swIn.center, circle, fill=red, inner sep=0pt,minimum size=1mm, anchor=center] {};
  \draw [thick, draw=red]($(swInput)+(0,0.15)$)--(swU.center);

  \node (swD) [below = 0.15cm of swIn.center, circle, fill=red, inner sep=0pt,minimum size=1mm, anchor=center] {};
  \draw [thick, draw=red]($(swInput)-(0,0.15)$)--(swD.center);

  \coordinate [right = -0.2cm of switch]  (swOut);
  \node (swO) [below = 0.0cm of swOut.center, circle, fill=red, inner sep=0pt,minimum size=1mm, anchor=center] {};
  \draw [thick, draw=red](switch.east)--(swO.center);
  \draw [thick, draw=red](swU.center)--(swO.center);

  \node(AI)[above = 2cm of controller, rectangle, minimum width = 4.0cm, minimum height = 0.8cm] {};

  \node(recordedData)[draw=red, ultra thick, left = 0.5cm of AI, rectangle, minimum width = 3.0cm, minimum height = 0.8cm] {\color{red}{\begin{tabular}{c}\textbf{recorded process}\\ \textbf{data}
  \end{tabular}
  }};

  \draw [-{Latex[length=2mm, width=2mm]}, draw=red, ultra thick] (recordedData.east)-|($(controller.north)-(0.15,0)$);

  \draw [-{Latex[length=2mm, width=2mm]}, draw=red, ultra thick, dashed] ($(disturbances.west)+(0.4,0)$)-|($(controller.north)+(0.15,0)$)node [above, pos=0.25] {\color{red}{\textbf{prediction}}};

  \end{tikzpicture}
  }
  }

  \subfloat[AI-based controller with a limited impact on the control values \label{fig:CLS_safety_b}]  {
  \resizebox{\linewidth}{!}
  {
  \begin{tikzpicture}
  \node(controller)[draw, rectangle, minimum width = 2.35cm, minimum height = 0.8cm] {\textbf{controller}};

  \node(parameter)[above = 0.3cm of controller, rectangle, minimum height = 0.8cm] {\begin{tabular}{c}
  static parameters \\ \footnotesize $(K_{P},K_{I},K_{D},\ldots)$
  \end{tabular}
  };

  \node(sum2)[draw=red, ultra thick, right = 1.25cm of controller, circle, minimum height = 0.8cm] {
  };
  \draw [draw=red, ultra thick] (sum2.north east) -- (sum2.south west)
  (sum2.north west) -- (sum2.south east);

  \node(sumtext)[below = 0.0cm of sum2, rectangle, minimum height = 0.8cm] {\color{red}\begin{tabular}{c}
  fine tune  \\ control variable \\  in predefined \\  boundaries
  \end{tabular}
  };

  \node(process)[draw, right = 3.4cm of sum2, rectangle, minimum width = 4.0cm, minimum height = 0.8cm] {\textbf{process}};

  \node(disturbances)[above = 0.3cm of process, rectangle, minimum height = 0.8cm] {\begin{tabular}{c}
  disturbances \\ \footnotesize $(D_{(t)})$
  \end{tabular}
  };

  \node(sensor)[draw, below right = 2.75cm and 3.35cm of controller, rectangle, minimum width = 4.0cm, minimum height = 0.8cm, anchor=center] {\textbf{sensor}};

  \node(response)[right = 0.7cm of process, rectangle, minimum height = 0.8cm] {\begin{tabular}{c}
   response \\ \footnotesize $(X_{(t)})$
  \end{tabular}
  };
  \node(sum)[draw, left = 0.3cm of controller, circle, minimum height = 0.8cm] {
  };
  \draw [thick] (sum.north east) -- (sum.south west)
  (sum.north west) -- (sum.south east);

  \node(reference)[left = 0.5cm of sum, rectangle, minimum height = 0.8cm] {\begin{tabular}{c}
   reference \\ \footnotesize $(W_{(t)})$
  \end{tabular}
  };

  \node(controller2)[draw=red, above right = 0.4cm and 0.1cm of controller, ultra thick, rectangle, minimum width = 2.35cm, minimum height = 0.8cm,fill=white] {\color{red}{\textbf{AI}}};

  \draw [-{Latex[length=2mm, width=2mm]}, thick, draw=black](sum2.east)--(process.west) node [above,pos=0.5] (controlVariables) {control variables} node [below,pos=0.5] {\footnotesize $(X_{1(t)},X_{2(t)},X_{3(t)},\ldots)$};
  \draw [-{Latex[length=2mm, width=2mm]}, thick, draw=black](process.east)--($(response.west)+(0.3,0)$) node (rs) [circle, pos=0.3, fill=black, inner sep=0pt,minimum size=1mm] {};

  \draw [-{Latex[length=2mm, width=2mm]}, thick, draw=black]($(reference.east)+(-0.3,0)$)--(sum.west);
  \draw [-{Latex[length=2mm, width=2mm]}, thick, draw=black]($(disturbances.south)+(0,0.1)$)--(process.north);
  \draw [-{Latex[length=2mm, width=2mm]}, thick, draw=black]($(parameter.south)+(0,0.1)$)--(controller.north);
  \draw [thick, draw=black](rs.center)|-(sensor.east);

  \draw [-{Latex[length=2mm, width=2mm]}, thick, draw=black]($(sensor.west)-(0,0.15)$)-|(sum.south) node [right,pos=0.95] {\textbf{--}};
  \draw [-{Latex[length=2mm, width=2mm]}, ultra thick, draw=red]($(sensor.west)+(0,0.15)$)-|($(controller2.south)-(0.8,0)$);

  \node (rsw) [right = 0.2cm of controller, circle, fill=red, inner sep=0pt,minimum size=1.5mm, anchor=center] {};

  \draw [-{Latex[length=2mm, width=2mm]}, ultra thick, draw=red] let
  \p1=($(rsw.center)$), \p2=($(controller2.south)$) in (rsw.center)--(\x1,\y2);

  \draw [-{Latex[length=2mm, width=2mm]}, ultra thick, draw=red] let
  \p1=($(sum2.north)$), \p2=($(controller2.south)$) in (\x1,\y2)--(sum2.north)node [right,pos=0.5] {\color{red}\textbf{+}};

  \draw [-{Latex[length=2mm, width=2mm]}, thick, draw=black](sum.east)--(controller.west);
  \draw [-{Latex[length=2mm, width=2mm]}, ultra thick, draw=red](controller.east)--(sum2.west) node [above,pos=0.85] {\color{red}\textbf{+}};;

  \node(AI)[above = 2cm of controller, rectangle, minimum width = 4.0cm, minimum height = 0.8cm] {};

  \node(recordedData)[draw=red, ultra thick, left = 0.5cm of AI, rectangle, minimum width = 3.0cm, minimum height = 0.8cm] {\color{red}{\begin{tabular}{c}\textbf{recorded process}\\ \textbf{data}
  \end{tabular}
  }};

  \draw [-{Latex[length=2mm, width=2mm]}, draw=red, ultra thick] (recordedData.east)-|($(controller2.north)$);

  \draw [-{Latex[length=2mm, width=2mm]}, draw=red, ultra thick, dashed] ($(disturbances.west)+(0.4,0)$)--($(controller2.east)$)node [above, pos=0.5] {\color{red}{\textbf{prediction}}};

  \end{tikzpicture}
  }
  }
\caption{Controlling safety critical processes by AI-based controller without the need of explainable AI.}
\label{fig:CLS_safety}
\end{figure}
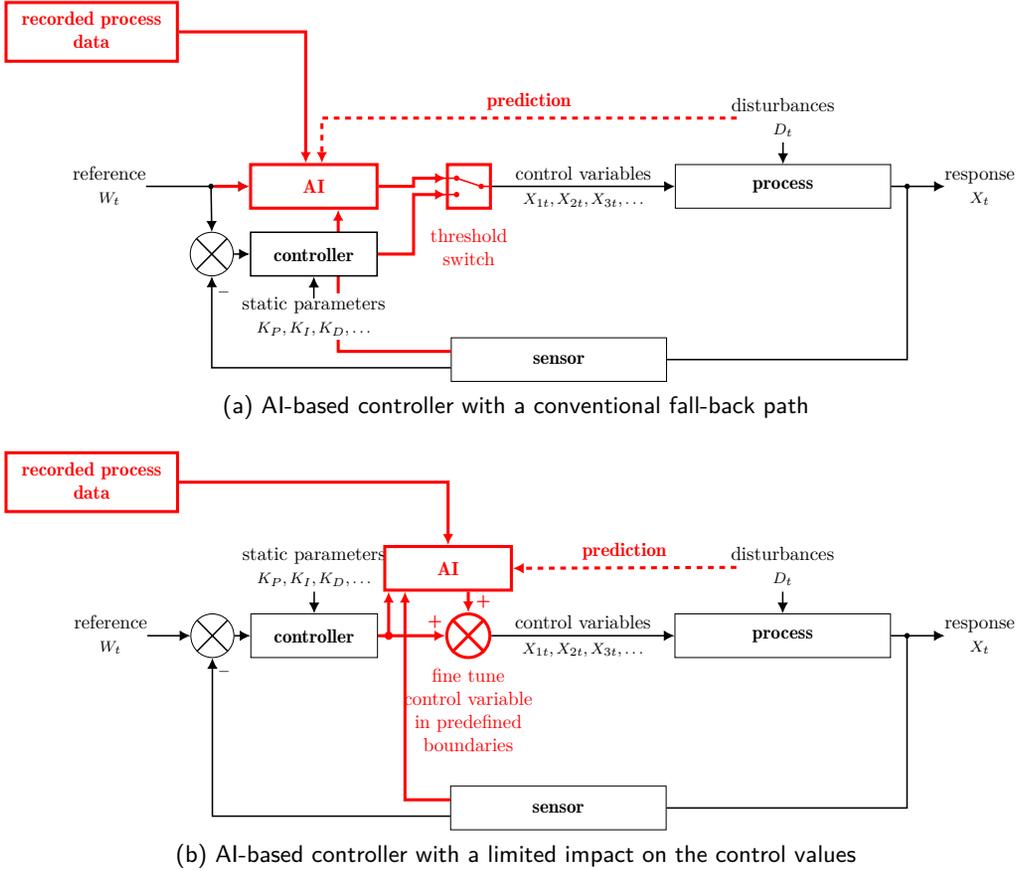

Since explainable AI, i.e., explainable deep ANN, does not exist yet \cite{Ras2020}, AI-based controller seems no option for controlling processes. Enabling the benefits of AI-based controllers and countering the risks of CLCS impossible to evaluate and test, two possible designs are depicted in Figure \ref{fig:CLS_safety}. To guarantee functional safety, one strategy is to provide a fall-back path in combination with a safety switch, disabling the AI-based controller depending on, e.g., thresholds. The block diagram of Figure \ref{fig:CLS_safety} \subref{fig:CLS_safety_a} illustrates this design. Theoretically, implementing a fall-back path using a conventional controller seems to be trivial. Indeed, the non-trivial part is the parameterization of the switch deciding whether to pass the AI-based controller's or the fall-back controller's control variables. Further, to improve functional safety, the number of trainable weights needs to be reduced, so that works \cite{Richter2021} on tailoring ANN needs to be considered.

Avoiding a head cut between possible unsafe AI-based controllers and a conventional ones, a second strategy is realized in Figure \ref{fig:CLS_safety} \subref{fig:CLS_safety_b}. Here, a conventional controller provides the control values. The AI-based controller fine-tunes these values in predefined boundaries--usually by symmetrical boundaries by plus-minus a defined offset for each control value. Due to the predefined maximum influence of the AI-based controller on the control values, this CLCS design can be evaluated and tested to fulfill the functional safety requirements.


\section{Conclusion}\label{sec:futherSteps}
AI in the field of control system engineering will be exploited for creating empirical process models by ANN. AI-based process modeling from recorded process data will reduce modeling time drastically and allow the conventional without any drawback. However, recorded process data are needed to cover the variances without biases. Thus a research focus must be on how these data can be recorded systematically, e.g., by stimulating a mechatronic actuator in a particular manner.

The AI-empowered static parameter set tuning will also be a pretty promising attempt for using the benefits of AI shortly. Nevertheless, due to its safety constraints, online AI-based changes of the controller parameters are not beneficial for most processes yet. The same lack in functional safety have entirely AI-based controllers. However, giving these kinds of controllers a defined impact on the control values of the process, as shown in Figure \ref{fig:CLS_safety} \subref{fig:CLS_safety_b}, is a good starting point for studying the pros and cons of AI-based controllers within deployed CLCS.

Finally, AI in the control system domain offers the kind of momentum for new control systems design, reflecting decades of knowledge and combining these with the process information in the form of recorded data. AI will not replace the experiences needed for control system engineering but will increase the importance of systematic recording process data.





\bibliographystyle{ACM-Reference-Format}
\bibliography{bib}

\includegraphics[]{pngSeed}

\end{document}